\documentclass[twocolumn]{aastex61}

\usepackage{color}
\usepackage{booktabs}

\begin{document}

\title{MODELING THE EFFECTS OF INHOMOGENEOUS AEROSOLS ON \\ THE HOT JUPITER KEPLER-7\MakeLowercase{b}'S ATMOSPHERIC CIRCULATION }
\author{Michael Roman and Emily Rauscher}
\affiliation{Department of Astronomy, University of Michigan, Ann Arbor, MI 48013, USA}

\begin{abstract}
Motivated by the observational evidence of inhomogeneous clouds in exoplanetary atmospheres, we investigate how proposed simple cloud distributions can affect atmospheric circulations and infrared emission.  We simulated temperatures and winds for the hot Jupiter Kepler-7b using a three-dimensional atmospheric circulation model that included a simplified aerosol radiative transfer model.  We prescribed fixed cloud distributions and scattering properties based on results previously inferred from Kepler-7b optical phase curves, including inhomogeneous aerosols centered along the western terminator and hypothetical cases in which aerosols additionally extended across much of the planet's night side.  In all cases, a strong jet capable of advecting aerosols from a cooler nightside to dayside was found to persist, but only at the equator.  Colder temperatures at mid- and polar-latitudes might permit aerosol to form on the dayside without the need for advection.  By altering the deposition and redistribution of heat, aerosols along the western terminator produced an asymmetric heating that effectively shifts the hottest spot further east of the sub-stellar point than expected for a uniform distribution.  The addition of opaque high clouds on the nightside can partly mitigate this enhanced shift by retaining heat that contributes to warming west of the hotspot.  These expected differences in infrared phase curves could place constraints on proposed cloud distributions and their infrared opacities for brighter hot Jupiters. 
\end{abstract}

\section{INTRODUCTION}

Inferences of clouds or hazes in exoplanetary atmospheres have recently provided new context for investigating the effects of aerosols in novel conditions.  The extreme temperatures and dynamics found on some exoplanets are expected to yield aerosols with compositions and distributions unlike those known in our solar system \citep{Lodders2010, Marley2013exoclouds}.  Limited in both data and theory, our currently poor knowledge of exoplanetary aerosols challenges our ability to interpret observations and characterize cloudy atmospheres \citep{SingNature2016,Stevenson2017}.  

Indications of aerosols have come from their presumed effects in both attenuating and scattering radiation beyond what would be expected from a clear, gaseous atmosphere alone. Muted or flattened features in both visible and infrared transmission spectra have been attributed to the presence of attenuating aerosols \citep{Pont2013,SingNature2016}, as have unexpectedly high visual albedos \citep{Demory2011} and anomalously low infrared emission \citep{Stevenson2017}.  While transmission spectra may be the most promising in potentially determining the composition and vertical extent of the aerosols \citep[e.g.][]{WakefordSing2015}, observations of reflected and emitted radiation over complete orbital phases can provide the best clues to how aerosols are distributed spatially over the planet, which itself can be diagnostic of aerosol composition and formation. By comparison to expected emission or reflectance from a clear atmosphere, such phase curves have been used to suggest the presence of inhomogenous aerosols with preferences for a particular hemisphere. In the case of WASP-43b, thermal phase curves from \textit{Hubble Space Telescope} (WFC3) had indicated a strong day-night temperature contrast \citep{Stevenson2014}, subsequently corroborated by \textit{Spitzer Space Telescope} \citep{Stevenson2017}, that was inconsistent with the more modest day-night temperature contrast predicted by clear atmospheric circulation models \citep{Kataria2015}.  As one possible explanation for this discrepancy between the observed and predicted nightside fluxes, \cite{Kataria2015} cited previous studies (specifically \cite{Showman&Guillot2002}) that suggested high altitude aerosols may be blanketing the nightside, decreasing atmospheric transmission and reducing the infrared emission to space. Likewise, unexpected asymmetry in the \textit{Kepler Space Telescope} optical phase curve of Kepler-7b has been interpreted as evidence of a reflective cloud layer distributed only along the western edge of the disk, scattering incident stellar radiation and increasing the observed visible flux \citep{Demory2013,Hu2015, Webber2015}.  In both cases, it is easy to imagine that these are clues that clouds may preferentially exist on or along the nightsides of these planets, where temperatures may be cool enough for aerosols to form from condensation or some temperature-dependent chemistry.  Such clouds, hazes, or more generally, aerosols, would presumably play a role in the planet's climate, and questions of their effect on atmospheric circulation, energy balance, and observations motivate a need for investigation through numerical modeling. 

The challenge and computational expense of three-dimensionally modeling aerosols over the range of scales and physical processes involved has led to a diversity of approaches. The most complex treatments include self-consistent, spontaneous cloud formation with modeled cloud microphysics, gaseous depletion, and feedback effects on cloud advection, settling, and multi-spectral radiative transport \citep{Lee_kitchensink2016, Lee_dynamical_clouds2017}.  A simpler yet valuable approach has been to post-process general circulation model (GCM) output to determine where clouds may most likely exist based on the temperatures and condensations curves \citep{Kataria2016,Oreshenko2015,Parmentier2016}.  The diversity of atmospheric conditions and considerable uncertainties involved arguably warrant a range of approaches, using a variety of assumptions and levels of complexity to investigate different aspects of the atmospheric physics. In the present paper, we investigate the effects of inhomogeneous clouds on a hot Jupiter atmosphere using a newly modified version of the GCM of \cite{RauscherMenou2012} that includes double-gray radiative forcing by aerosols with fixed spatial distributions. The double-gray approximation significantly reduces the computational burden and has been successfully employed in previous modeling \citep[e.g.][]{Dobbs-Dixon2010,Heng2011,RauscherMenou2012}. Although it neglects the considerable wavelength-dependence of scattering across each spectral channel, the double gray approximation captures the essential physics of visible and thermal radiative transport necessary for computing heating rates to first order, while allowing us to investigate a larger parameter space more efficiently. Similarly, prescribing the aerosol distribution considerably reduces the complexity of aerosol modeling, allowing us to isolate the role of cloud spatial coverage from the many factors that would be necessary to more fully model the cloud physics \citep[e.g.][]{Lee_kitchensink2016,Lee_dynamical_clouds2017}.  This differs from the approach of post-processing the final GCM results since we include the radiative forcing from aerosols in each time step throughout the simulation, which serves to alter the temperature and wind fields of the developing model. Rather than use the clear atmospheric simulation to tell us where aerosols should be, we impose the aerosol locations and see how the temperature and winds evolve in their presence.  We consider this to be complementary to works that investigate specific aspects of aerosols \citep[e.g.]{HengDemory2013,Parmentier2013,Parmentier2016}, as well as to models of higher complexity in which many aspects of the physics influence each other in detailed ways \citep[e.g.][]{Lee_kitchensink2016}.

As a case study, we model the general circulation of Kepler-7b \citep{Latham2010}, a hot Jupiter for which specific, inhomogeneous aerosol distributions have been proposed to explain observations.  Measured occultation depths from \textit{Kepler} observations initially suggested Kepler-7b had a surprisingly high planetary albedo or brightness temperature \citep{Demory2011}.  Subsequent infrared \textit{Spitzer} observations placed upper limits on the thermal emission and refined the albedo \citep{Demory2013}.  Combined with the infrared constraints, analysis of the phase curves showed an asymmetry in the reflected light over an orbital period.  This was interpreted as possible evidence of an asymmetry in the spatial coverage of clouds, specifically suggesting a reflective cloud layer distributed along the western edge of the disk. \cite{MnI2015} subsequently fitted models of aerosol distributions and scattering parameters by calculating phase curves produced by a layer of multiple-scattering spherical particles suspended above a low-albedo surface (representing the gaseous atmosphere). Degeneracies between the cumulative aerosol optical thickness, the assumed albedo of the atmosphere beneath the aerosol layer, and precise central longitude of the distribution permitted a few different combinations of parameters.  In all cases, a high, very optically thick layer of conservative scatterers centered near the western limb was found necessary to produce the asymmetry in the optical phase curve.  

Three-dimensional atmospheric models of tidally locked hot Jupiters consistently display broad, prograde equatorial jets \citep[e.g.][]{showman2009atmospheric, Dobbs-Dixon2010, RauscherMenou2010, Thrastarson&Cho2010, Heng2011, Mayne2014} ultimately driven by strong day-night thermal forcing \citep{ShowmanPolvani2011}. Indeed, such a circulation pattern has been reasonably invoked to explain the hemispheric asymmetry in the aerosols suggested by the Kepler-7b optical phase curve \citep{MnI2015}.  Clouds may form near or beyond the western terminator, where cooler gas advected from the nightside may support condensation \citep{Parmentier2016}.  The aerosols are then advected by eastward winds to the strongly-irradiated dayside, where the aerosols are no longer in thermo-chemical equilibrium.  They presumably survive just long enough to be seen along the western terminator before returning to the gaseous state.  

The presence of an extensive, optically thick, highly reflective aerosol layer would not only reduce the global bond albedo of Kepler-7b, but would alter the pattern of local radiative heating and consequent wind field.  This prompts the questions of how such a cloud would alter the atmospheric circulation, and whether or not the resulting pattern would remain self-consistent with the proposed advection mechanism.  Additionally, how might this proposed cloud affect the thermal emission and consequent infrared phase curves?

Motivated by these questions, we first introduce our newly modified GCM with a simple aerosol model and discuss the model set-up for the simulations presented here (Section 2).  Using Kepler-7b as a case study, we then investigate how a prescribed inhomogeneous cloud distribution may affect the general circulation of an hot-Jupiter atmosphere (Section 3) with observational implications (Section 3.3). We discuss the consistency, or lack thereof, between our assumed aerosol distributions and actualized results of our GCM simulations (Section 4), and conclude with a summary of our main findings (Section 5).    

\section{ATMOSPHERE MODELING WITH INHOMOGENEOUS CLOUDS}

To investigate the effects of inhomogeneous cloud cover on a hot Jupiter atmosphere, we model the general circulation of Kepler-7b for a variety of proposed aerosol distributions consistent with the Kepler phase curve observations \citep{Demory2013} and phase curve modeling of \cite{MnI2015}.  Our modeling uses a newly modified version of the GCM of \cite{RauscherMenou2012}. Originally based on the Intermediate General Circulation Model of the University of Reading \citep{Hoskins1975}, the code had been adapted for modeling gaseous exoplanetary atmospheres, including hot Jupiters \citep{MenouRauscher2009, RauscherMenou2010,RauscherMenou2012}, a hot Neptune \citep{Menou2012a} and circumbinary Neptunes \citep{May2016}. The present version utilizes the same dynamical core to solve the primitive equations, but features a revised radiative transfer scheme to include the effects of aerosols following \cite{Toon1989}. General parameters defining the dynamics, radiative transfer, and spatial resolution of our model are listed in Table 1. We ran simulations employing a range of plausible aerosol distributions and scattering properties, as described below and listed in Table 2, and we compared the results to investigate the effect of the proposed aerosols on the temperature and wind fields. Although the atmosphere of Kepler-7b is hot enough that we should expect thermal ionization and the potential influence of magnetic effects \citep{Perna2010, Menou2012b, RauscherMenou2013, Rogers2014}, we neglect them from this modeling work in order to isolate the role of inhomogeneous aerosols in altering the atmospheric circulation.

\begin{deluxetable*}{lccc}[th!]
 \centering
\tabletypesize{\footnotesize}
\tablecolumns{4} 
\tablewidth{4in}
 \tablecaption{ General Model Parameters
 \label{tab:lumresults1}}
 \tablehead{
 \colhead{Parameter} & \colhead{Value} & \colhead{Units} & \colhead{Comment} }
 \startdata 
\it{    Orbital/Dynamical}\\
Radius of the planet, $R_p$ & $1.128\times 10^8$ & m & ref: Demory et al. 2011 \\
Gravitational acceleration, $g$ & 4.17 & m s$^{-2}$ & ...\\
Rotation rate, $\Omega$ & $1.49 \times 10^{-5}$ & s$^{-1}$ & assumed tidally synchronized\\\\
\it{    Clear Atmosphere Radiative Transfer}\\
Specific gas constant, $\mathcal{R}$ & 3523 & J kg$^{-1}$ K$^{-1}$ & assumed $H_2$ rich\\
Ratio of gas constant to heat capacity, $\mathcal{R}/c_P$ & 0.286 & -- & assumed Diatomic \\
Incident flux at substellar point, $F_{\downarrow \mathrm{vis}, \mathrm{irr}}$ & $1.589 \times 10^6$  & W m$^{-2}$ & ref: Demory et al. 2011\\
Internal heat flux, $F_{\uparrow \mathrm{IR}, \mathrm{int}}$ & 2325 & W m$^{-2}$  & from modeled T-profile\\
Visible absorption coefficient, $\kappa_{\mathrm{vis}}$  & $1.57 \times 10^{-3}$  & cm$^2$ g$^{-1}$ & constant, from modeled T-profile \\
Infrared absorption coefficient, $\kappa_{\mathrm{IR}}$  & $1.08 \times 10^{-2}$  & cm$^2$ g$^{-1}$  & ...\\
\it{     }Pressure of $\tau_{vis}$ = 1 for two-way vertical path & $88$ & mbar & gas only, calculated from $\kappa_{\mathrm{vis}}$ and $g$\\
\it{     }Pressure of $\tau_{vis}$ = 2/3 for one-way vertical path & $177$ & mbar & ... \\
\it{     }Pressure of $\tau_{IR}$ = 2/3 for one-way vertical path & $257$ & mbar & gas only, calculated from $\kappa_{\mathrm{IR}}$ and $g$ \\\\
\it{    Model Resolution}\\
Vertical layers & 50 & --\\
Bottom boundary pressure & 100 & Bar \\
Top boundary pressure & $5.7 \times 10^{-5}$ & Bar\\
Horizontal Resolution & T31 & -- & corresponds to 48 lat $\times$96 lon\\
Temporal Resolution & 4800 & time steps/day & \\
Simulated Time & 1000 & days & \\
 \enddata
 \vspace{-0.8cm}
\end{deluxetable*}

\subsection{Radiative Transfer Scheme} 

Aerosols have the potential to scatter and absorb radiation, thus directly altering the radiative transfer and consequent heating rates within the atmosphere.  To include these effects, we adopted the radiative transfer scheme of \cite{Toon1989}\footnote{There is a typographical error in Eq.(42) of \cite{Toon1989}.  We used the corrected form: 
\begin{eqnarray}
E_l=[C^+_{n+1}(0)-C^+_{n}(\tau_n)] e_{2n+1} + [C^-_{n}(\tau_n)-C^-_{n+1}(0)] e_{4n+1}\nonumber
\end{eqnarray}
}, which computes radiative heating rates for an inhomogeneous multiple-scattering atmosphere with a two-stream approximation. This replaced the previous scheme used in the code of \cite{RauscherMenou2012} that had solved the equations of transfer for purely absorbing atmospheres. All our calculations, for both clear and cloudy conditions, now utilize the newly adopted scheme.

A complete description and derivation of the radiative transfer scheme can be found in \cite{Toon1989}, but we provide a brief summary as follows.  The true radiation field is approximated by fluxes in two directions\textemdash{an upward stream and a downward stream.}  The atmosphere is divided into a series of homogenous layers for which temperatures and fluxes are desired. Parameters characterizing the scattering, absorption, and emission within each layer are prescribed as detailed in Section 2.2.2.  Boundary conditions are applied to the system by imposing a downward optical flux at the top of model, an upward thermal flux at the bottom, and flux continuity across each layer interface. This results in a system of flux equations which can then be arranged into a matrix and inverted to solve for the net fluxes at each layer interface. The divergence in net flux across each layer is finally used to compute the heating rates:

\begin{equation}
\frac{dT}{dt} \sim \frac{dF_{net}}{dp}\frac{1}{c_p}
\end{equation}

where, for each layer, \textit{T} is the temperature, \textit{$F_{net}$} is the net flux, \textit{$c_p$} is the gaseous heat capacity at constant pressure, and \textit{d/dt} and \textit{d/dp} are the derivatives with respect to time and pressure, respectively.
 
The radiative transfer for visible and infrared (IR) radiation are treated separately.  Following \cite{RauscherMenou2012} we retain the double-gray spectral approximation, which offers advantages in simplicity and computational efficiency at the cost of limiting the wavelength-dependent physics. In the visible band, incoming stellar radiation is incident upon the upper boundary and may be absorbed and scattered by both gas and aerosols; a $\delta$-quadrature two-stream approximation is used as recommended by Toon et al (1989). In the IR, the primary source of radiation is isotropic blackbody emission from each layer, with additional contribution from upwelling radiation at the bottom boundary (i.e. internal heat source).  Each atmospheric layer is divided into two sub-layers to better resolve the emission as a function of optical depth across each layer. Given just one IR band, the emitted flux is evaluated using $\sigma$$T^4$/$\pi$, effectively integrated over all wavelengths. Gaseous scattering in the IR is ignored, though aerosol scattering is included. To compute the net IR fluxes, a $\delta$-hemispheric mean two-stream approximation is first used to determine coefficients for the system of equations, followed by application of the source function technique of \cite{Toon1989}. This additional refinement improves the accuracy in the case of a purely absorbing atmosphere and involves using the initial two-stream solution to the radiation intensity as an approximation of the true intensity source function within the equation of transfer. For details of this approach, we direct the reader to \cite{Toon1989}.  Finally, as in the previous version of the code, we retain the dry convection adjustment for convectively unstable layers and an optional transition to flux-limited diffusion at sufficiently high optical depths \citep{RauscherMenou2012}.  

Gaseous absorption coefficients for the visible and IR channels are estimated from the expected atmospheric temperature profiles of Kepler-7b.  One-dimensional, cloud-free, dayside temperature profiles, modeled by \cite{Demory2013} using methods of \cite{Fortney2008unified}, were compared to solutions from simple analytical expressions of \cite{guillot2010radiative}. The absorption coefficients and the interior upwelling flux were then chosen from those analytical expressions that had best matched the 1D modeling.  We used a constant constant absorption coefficient, neglecting the effects of pressure broadening and collision-induced absorption that are important at higher pressures \citep{guillot2010radiative}.  We found this matched the more detailed 1D model well. Values for the absorption coefficients are included in Table 1.

\subsection{Aerosol Distribution and Scattering Parameters}
To ensure consistency with the \textit{Kepler} optical phase curve observations and significantly reduce the complexity of the modeling, the aerosol distribution was chosen to be prescribed and static.  We omit any time-dependence that would result from aerosol microphysics or advection. This simplification essentially assumes a steady-state equilibrium between cloud formation and dissipation at all times\textemdash {even as the model temperature and wind fields are still evolving.  The aerosol distribution is therefore in no way required to be self-consistent with the atmospheric environment, but this provides the advantage of isolating how the distribution of cloud cover can actively affect the atmospheric heating and circulation throughout the simulation.  Details of our aerosol models are discussed below and listed in Table 2.

\subsubsection{Spatial Distribution}

\cite{MnI2015} determined several solutions for aerosol distributions with reflectance consistent with Kepler-7b's observed optical light phase curve. Using a multiple-scattering model, they found that the asymmetry in the phase curve could be produced most simply with a horizontal, circular patch of clouds centered at the equator near the western terminator. The precise position and goodness-of-fit depended on the assumed reflectance of the background atmosphere, modeled as a Lambertian surface of reflectance \textit{ $r_g$}.  Accordingly, \cite{MnI2015} provided functional expressions for several distributions, along with corresponding aerosol scattering properties and $\chi$$^2$ evaluations, grouped into four cases, categorized by the value of the assumed background reflectance.  Their best fits were found assuming a background reflectance of 0.0 (reduced $\chi$$^2$~1.013) and 0.1 (reduced $\chi$$^2$~1.009). For our modeling, we adopt these two best-fitting cloud distributions and background reflectances and develop corresponding cloud models for each. Additionally, for comparison, we ran a model that included uniform, global cloud coverage.   In this case, the aerosol optical thickness was carefully chosen to produce a spherical albedo and a global energy balance equivalent to the inhomogeneous cloud case.  The resulting reflectance is therefore consistent with the global albedo of Kepler-7b, but phase curve would lack the observed asymmetry.  As a purely theoretical case, this global models allows us to compare and further isolate the effects due to aerosol inhomogeneity.

For simplicity, the fractional spatial coverage of aerosols within each cloudy grid cell (i.e. the cloud-fraction) is taken to be unity.  This implies a thick uniform layer of aerosols as opposed to a broken coverage.  Such overcast conditions would likely be unrealistic for an extensive condensate cloud, though may be more appropriate for assumptions of a thick, extended haze layer. This assumption maximizes the aerosol radiative forcing within each cloudy grid cell and can thus be considered a limiting case for the distribution. Future work will explore modeling with a range of cloud fractions.

Although the interpretations of the Kepler optical phase curves can only suggest an aerosol distribution for the visible dayside of Kepler-7b, it is reasonable to expect that clouds may also be present on the nightside.  Indeed, this would be physically consistent with a presumed scenario in which clouds form on the nightside, where cooler temperatures possibly permit condensation or supportive chemistry, before being advected and consequently vaporized on the hot dayside.  For the basic, circular distribution suggested by \cite{MnI2015}, we allow part of the cloud to extend over into the nightside rather than abruptly truncating it at the terminator.  To explore this possibility further, we also run models in which the cloud distributions at the western terminator are extended across most of the nightside of the planet, reaching the eastern terminator.  Regions approaching the eastern terminator are reduced in cloud opacity for self-consistency, since warm, clearer skies along the eastern terminator may presumably be advected to the night side by the same winds that carry aerosols to the dayside (see Figure \ref{fig:tfieldclouds}).  

The vertical distribution of aerosols was not constrained by \cite{MnI2015}, though the authors suggested the lack of background atmospheric reflectance (i.e. high atmospheric absorption) was consistent with very high clouds (above 10$^{-4}$ bar).  Given the inferred opacity of the atmosphere at visible wavelengths, the clouds would indeed need to be above $\sim$ 80 mbar in order to contribute significantly to the observed reflectance at even normal (substellar) incidence angles; towards the limb, the extended optical path would require aerosols to be even higher. These low pressures are also consistent with the expected condensation pressures of various magnesium silicates given the inferred temperature profile \citep{Demory2013}. Considering these constraints, we chose to place the base of our aerosol layer at 10 mbar. To produce the large optical thickness suggested by the data, we let the layer extend vertically to the top of the atmospheric model at 57 $\mu$bar (over 18 layers), falling in optical thickness at a rate equal to the drop in atmospheric pressure to maintain a constant mixing ratio (see Fig 3). This amounts to an aerosol abundance at the top of the model that is approximately only 1.25\% of its maximum value found at the base of the aerosol layer.  Even so, the slant path two-way transmission near the limb, through the thickest part of the cloud, becomes optically thick within the top layer due to the high stellar incidence angle.

Comparing our assumed vertical profile to the vertical distributions determined by \cite{Parmentier2013} from tracers in a GCM, our simple distribution compares reasonably well with the modeled distributions for 0.5-1$\mu$m particles.  \cite{Parmentier2013} shows that larger particles would tend to drop off more quickly, while 0.1$\mu$m particles would drop in abundance even more slowly with height. Different assumptions regarding the vertical mixing and aerosol sedimentation would lead to different vertical profiles, some quite complicated if multiple layers are formed \citep[e.g.][]{Lee_kitchensink2016, WakefordSing2015} or if aerosols extend deeper into the unobserved atmosphere; however, the precise choice of vertical profile does not significantly affect our results.  As a test, we ran simulations with clouds extending to from 10 mbar to 1mbar, 100$\mu$bar, and 57$\mu$bar and found all the results to be qualitatively very similar, with most absolute differences in temperatures, winds, and emitted fluxes of less than few percent. Evidently, provided that the aerosols are visible in the stable layers of the upper atmosphere, the horizontal differential heating will dominate the flow regardless of the aerosols' precise vertical distribution.

\subsubsection{Scattering Parameters}

In our modeling, the aerosol scattering for each wavelength channel is characterized by three parameters: the aerosol optical depth, the single scattering albedo, and the asymmetry parameter. The aerosol optical depth $\tau$ quantifies the attenuation of radiation due to the total aerosol component of an atmospheric layer; the single scattering albedo $\varpi_{0}$ defines the fraction of incident light scattered by each particle, with values ranging between 1 (conservative scattering) and 0 (fully absorbing); and the asymmetry parameter \textit{$g_0$} specifies the preferential direction of aerosol scattering representative of each particle's scattering phase function, with values again ranging between 1 (strongly forward scattering), 0 (isotropically scattering), and -1 (strongly backward scattering). These scattering parameters can theoretically be related to the particle's physical characteristics, including the size, shape, and composition, though typically non-uniquely.

Based on the extremely high reflectance and weak forward scattering necessary to match the optical phase curves, \cite{MnI2015} concluded that the clouds of Kepler-7b were optically very thick and composed of tiny particles (0.1-0.4 $\mu$m in radius).  The total integrated optical thickness and scattering asymmetry parameter were dependent on the cloud distribution and assumed underlying atmospheric albedo.  In all cases, the inferred aerosol single scattering albedos were $\varpi$$_0$ $\approx$ 1.  Best fitting asymmetry parameters were found to be between 0.4 and 0.5. From these optical values, the authors inferred possible candidates for the particle composition given the assumed standard atmospheric composition and expected temperatures ($\approx$1,700 K at the relevant heights).  These included two silicates (Mg$_2$SiO$_4$ and MgSiO$_3$), perovskite (CaTiO$_3$), and silica (SiO$_2$). For our GCM simulations, we chose scattering parameters guided by the \cite{MnI2015} results.  

The single scattering albedo and asymmetry parameters of the aerosols in the visible channel were adopted directly from \cite{MnI2015}; however, we found that their total integrated optical depths were insufficient in reproducing their derived spherical albedos within our model.  That our model should require additional aerosol to produce the same reflectance is not surprising given the fundamental differences in modeling; \cite{MnI2015} employed a rigorous multiple-scattering model that consisted of a cloud of spherical particles suspended above a Lambertian layer; in our model, we use a simple two-stream approximation with aerosols interspersed within the absorbing gas. In order to obtain a similar energy balance in the visible channel, we increased our aerosol optical depths by roughly 60-70\% until our globally averaged, visible top-of-the-atmosphere albedos matched the quoted spherical albedos of \cite{MnI2015} for each case. 

\begin{deluxetable*}{lccc}[h]
\tabletypesize{\footnotesize}
\tablecolumns{4} 
\tablewidth{0pt}
 \tablecaption{ Aerosol Model Parameters}
 \label{tab:lumresults1}
 \tablehead{\colhead{Parameter} & \colhead{Value} & \colhead{Comment}}
 \startdata 
 \it{Assumed Atmospheric background albedo = 0.0} & &following M\&I(2015), rg = 0\\
{       }{ }Maximum visible aerosol optical depth, $\tau$$_{c,vis}$ & 159 & yields A$_s$ $\sim$ 0.455\\
{       }{ }Global model visible aerosol optical depth, $\tau$$_{c,vis}$ & 1.8 & yields A$_s$ $\sim$ 0.455\\
{       }{ }Central longitude offset, $\Delta$$\phi$$_c$ & -65$^{\circ}$& longitude of  $\tau$$_c$ peak, east from sub-stellar\\
{       }{ }Horizontal extent parameter, $\sigma$$_c$ & 25$^{\circ}$ & $\tau$$_c$ falls off as exp(1/2$\sigma$$_c$$^{2}$) from center\\
{       }{ }Visible single scattering albedo, $\varpi_{0,vis}$ & $\sim$ 1.00& following M\&I(2015)\\
{       }{ }Visible scattering asymmetry parameter, $g_{0,vis}$ & 0.376& ...\\
{       }{ }Maximum IR aerosol optical depth, $\tau$$_{c,ir}$ &0.95, 3.2 &based on Silica$^\dagger$, Perovskite$^\ddagger$ at 5$\mu$m\\
{       }{ }IR scattering asymmetry parameter,$g_{0,ir}$&  0.006, 0.09  & ...\\
{       }{ }IR single scattering albedo,$\varpi_{0,ir}$   &  0.045, 0.95&... \\\\
\it{Assumed Atmospheric background albedo = 0.1} &&following M\&I(2015), rg = 0.1\\
{      }{ }Maximum visible aerosol optical depth, $\tau$$_{c,vis}$ & 254 & yields A$_s$ $\sim$ 0.461\\
{      }{ }Central longitude offset, $\Delta$$\phi$$_c$ & -75$^{\circ}$& longitude of  $\tau$$_c$ peak, east from sub-stellar\\
{      }{ }Horizontal extent parameter, $\sigma$$_c$ & 25$^{\circ}$ & $\tau$$_c$ falls off as exp(1/2$\sigma$$_c$$^{2}$) from center\\
{      }{ }Visible single scattering albedo, $\varpi_{0,vis}$ & $\sim$ 1.00 & following M\&I(2015)\\
{      }{ }Visible scattering asymmetry parameter, $g_{0,vis}$ & 0.416 & ...\\
{      }{ }Maximum IR aerosol optical depth, $\tau$$_{c,ir}$ &0.6, 1.2  & Silica$^\dagger$ at 5$\mu$m, Perovskite$^\ddagger$ at 10$\mu$m\\ 
{      }{ }IR scattering asymmetry parameter,$g_{0,ir}$& 0.007, 0.02  & ...\\
{      }{ }IR single scattering albedo,$\varpi_{0,ir}$   & 0.06, 0.78 &... \\\\
\enddata

 \tablecomments{The expression for the aerosol distribution comes from \cite{MnI2015}:\\
$\tau(\phi,\psi;\tau_c,\sigma_c,\Delta\phi_c$) = $\tau_c$ exp [ --(\{$\phi - \Delta\phi_c\}^{2} + \psi^{2}) / (2\sigma_c^{2})]$, where $\phi$ and $\psi$ are the longitude and latitude, respectively; other parameters are as defined above.\\
$^\dagger$\cite{kitamura2007optical}, $^\ddagger$\cite{posch2003infrared} $^\ddagger$\cite{WakefordSing2015},$^\ddagger$\cite{zeidler2011near}}
\end{deluxetable*}

With our double-gray modeling, additional choices were required for aerosol scattering parameters at infrared wavelengths, beyond those directly implied by visible observations. The most important parameter is the extinction efficiency of the IR scatterers relative to the extinction at visible wavelengths, which determines the infrared opacity of the clouds. Given a roughly spherical particle of a known size and complex index of refraction, the basic scattering parameters at any wavelength may be estimated using Mie theory.  The particle sizes and compositions deduced from the \textit{Kepler} optical observations can thereby be used to guide choices for self-consistent scattering properties at longer wavelengths.

Of the minerals suggested, silica and perovskite have the respective lowest and highest real refractive indices, and thus together yield a broader range of scattering parameters. Hence, we chose to limit our Mie modeling of infrared scattering parameters to these two compositions.  For particles with roughly 0.1-0.4$\mu$m radii, both minerals scatter isotropically at infrared wavelengths, but the perovskite produces a higher single scattering albedo and a roughly threefold greater extinction efficiency compared to silica.  However, even with the composition and particle size inferred, choosing a wavelength at which to evaluate the IR scattering parameters is not obvious in context of our double-gray approach.  The infrared channel attempts to capture the essential physics of infrared radiative transfer, but it does not define a precise range or central value.  Over a broad range in frequencies, the cloud opacity would presumably range from optically thick in the near- and mid-IR to optically very thin at far-infrared wavelengths.  So though an aerosol layer may be optically thick to thermal emission at 1.7$\mu$m (the blackbody peak wavelength corresponding to 1,700K), a considerable amount of heat may still radiatively escape beyond a few microns, where particles attenuate less.  With these considerations, we chose to run simulations with a range of aerosol infrared opacities by evaluating silica at 5$\mu$m and perovskite at both 5$\mu$m and 10$\mu$m.  Given the assumed optical opacities, this leads to maximum vertically integrated IR aerosol opacities that range from 0.6 to 3.2, i.e. optically thin to optically thick.  The chosen aerosol parameters in the both bands are summarized in Table 2. 

A gaseous scattering was also introduced to provide an additional source of reflection, as when needed to match the clear atmosphere albedo \cite{MnI2015} results given differences in our modeling approaches.  As noted, that study had used a Lambertian background layer to simulate the atmospheric reflectance coming from beneath the aerosol layer.  For our modeling, we chose to reproduce this reflectance by treating it as Rayleigh scattering emerging from the gaseous atmosphere.  Rayleigh scattering cross-sections as a function of wavelength were first computed for our assumed hydrogen-rich atmosphere (consistent with our assumed gas constants). We then evaluated the Rayleigh scattering optical depth at precisely the wavelength required to match the spherical albedo of the background atmosphere as required by \cite{MnI2015}.  For matching their case of a purely absorbing background atmosphere, no Rayleigh scattering was included.  For cases in which the background albedo was equal to 0.1, we found that evaluating the Rayleigh scattering at 410nm (in combination with our inferred visible absorption coefficient) provided the correct reflectance.  This frequency happens to fall just beyond the Kepler passband (5\% at 423-897 nm, Koch et al. 2010), but such albedo could hypothetically still be achieved in a gray approximation by tweaking the gaseous composition or introducing an additional uniform aerosol layer.  We note that for a spectrally resolved band, \cite{Demory2011} found that Rayleigh scattering alone could produce sufficiently high albedos in the Kepler passband if the strongest absorbers of Rayleigh scattering---Na and K--- were depleted by a factor of 10-100 of the equilibrium composition. For our modeling, we simply let the Rayleigh optical provide all the clear atmospheric reflectance.  

\subsection{Summary of Model and Parameter Choices}
To summarize, eleven simulations were run in total.  Eight cloud simulations were run to cover a range of scattering solutions consistent with those proposed by \cite{MnI2015}, with their two different albedo assumptions, and our different choices for the infrared scattering properties (silica vs. perovskite) and basic cloud distribution (western terminator vs. western terminator+nightside).  The remaining three simulations included a uniform, global cloud model with optical depths chosen to match the spherical albedos of the inhomogeneous cloud models, a clear case with expected Rayleigh scattering, and a clear case with no scattering. Parameter values are listed in Table 2.

The model was initiated with the analytical temperature profiles and still wind at all locations.  Simulations were run for 1000 days of model time at temporal resolution of 4800 time steps per day. To evaluate whether a quasi-steady state solution had been reached, we evaluated the kinetic energy as a function of height and time as done in \cite{RauscherMenou2010}.  Typically, above about 7 bars, the kinetic energy reached a near constant value; below these heights, at pressures much deeper than the levels we observe, the kinetic energy continued to slowly increase.  From this we concluded that 1000 days of simulation time was long enough for the atmosphere to settle into a steady state at the pressures we observe. Different spatial resolutions were investigated to ensure that we captured the general circulation.  No significant changes were evident for resolutions greater than our chosen spectral model grid resolutions of T31, which corresponds to grid of 48 latitudes by 96 longitudes. Model resolutions are summarized in Table 1.

\section{RESULTS}

We compared expected wind and temperature fields derived from our general circulation model for twelve different models, ranging from clear to largely cloudy. As with previous modeling of clear hot Jupiters \citep{showman2009atmospheric,Dobbs-Dixon2010,RauscherMenou2010,Heng2011,Mayne2014}, at IR photospheric pressures, we found the hottest locations advected to the east of the substellar point due to a strong eastward jet.  This was seen in all cases, regardless of cloud model, implying that the standard dynamical pattern we have come to expect for hot Jupiters is robust against deviations from the strictly day-night hemispheric forcing pattern analyzed in the definitive work by \cite{ShowmanPolvani2011}. We have already seen the dynamical pattern is maintained for most, but not all, non-synchronous rotational rates tested \citep{showman2009atmospheric, RauscherKempton2014}. In the models presented here, we see that even when significantly altering the dayside heating pattern by strongly obscuring part of the hemisphere, the dynamics still produce the standard eastward equatorial jet.  This persistent pattern, however, was modulated by the choice of aerosol model, as discussed below, with potential observational implications.  

Though results differed slightly based on the scattering parameters in each aerosol model, we found we could reasonably group the results into four characteristically different cases, as follows: 

\begin{itemize}
    \item Models with no scattering or just Rayleigh scattering, which we will refer to as \textit{clear} cases
    \item Our \textit{global cloud} case, in which a cloud of optical depth 1.8 covers all locations with global spherical albedo equivalent to the inhomogeneous cloud cases
%    \end{itemize} 
%\item Inhomogeneous Cloud Models: 
%\begin{itemize} 
    \item Models with aerosols centered near the western terminator, hereafter referred to as \textit{western terminator cloud} cases
    \item Models with aerosols centered on the western terminator plus much of the nightside, hereafter referred to as \textit{western terminator+nightside cloud} cases
%\end{itemize}
\end{itemize}

Representative examples of results from each these cases are shown here, and differences within each case are discussed when significant.  The same four cases are shown in all the results.  These cases use the parameters listed in Table 2;  specifically, they assume zero reflectance from the clear atmosphere (i.e. rg=0) and scattering parameters for perovskite in the thermal (i.e. $\tau_{c,ir}$=3.2, $g_{0,ir}$=0.09, and $\varpi_{0,ir}$=0.95).  Overall, we did not find any significant qualitative differences between results computed using the two different albedo assumptions and corresponding scattering parameters of \cite{MnI2015}, but the choice of infrared scattering parameters had notable impact on the western terminator+nightside cloud cases as discussed below. For completeness, results for all scattering assumptions are included in Appendix \ref{AppB}.

\subsection{Winds}

\begin{figure*}[t]
\includegraphics[width=\textwidth]{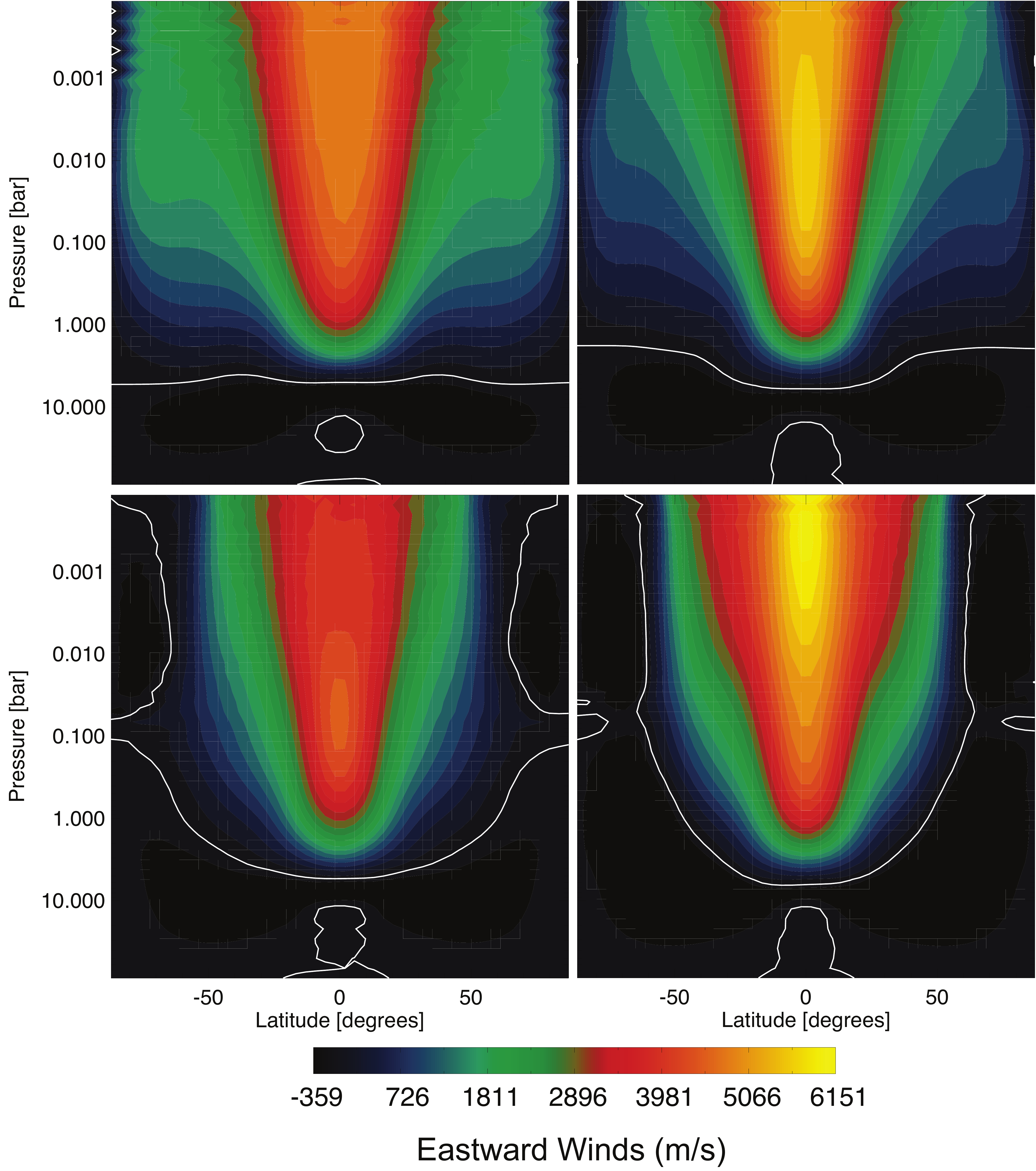}
\caption{Zonally averaged wind cross sections four representative cases: clear, purely absorbing atmosphere (upper left), global cloud (upper right), western terminator cloud centered 65$^{\circ}$ W of the substellar point (lower left), and western terminator+nightside (lower right).  Winds are in units of meters per second, eastward positive. The zero contour is indicated by the white line.  Eastward winds are capable of advecting potential clouds and cooler gas from the night side across the western terminator. All models show a strong eastward jet along the equator, although the strength and width of this feature varies between different cases. These aerosol models use the following scattering parameters:  rg=0, $\tau_{c,vis}$=159, $g_{0,vis}$=0.376, $\varpi_{0,vis}$=1.0, $\tau_{c,ir}$=3.2, $g_{0,ir}$=0.09, and $\varpi_{0,ir}$=0.95.}
\label{fig:zonalwinds}
\end{figure*}
%%%%%%%%%%%%%%%%%%

\begin{figure*}[th!]
\includegraphics[width=\textwidth]{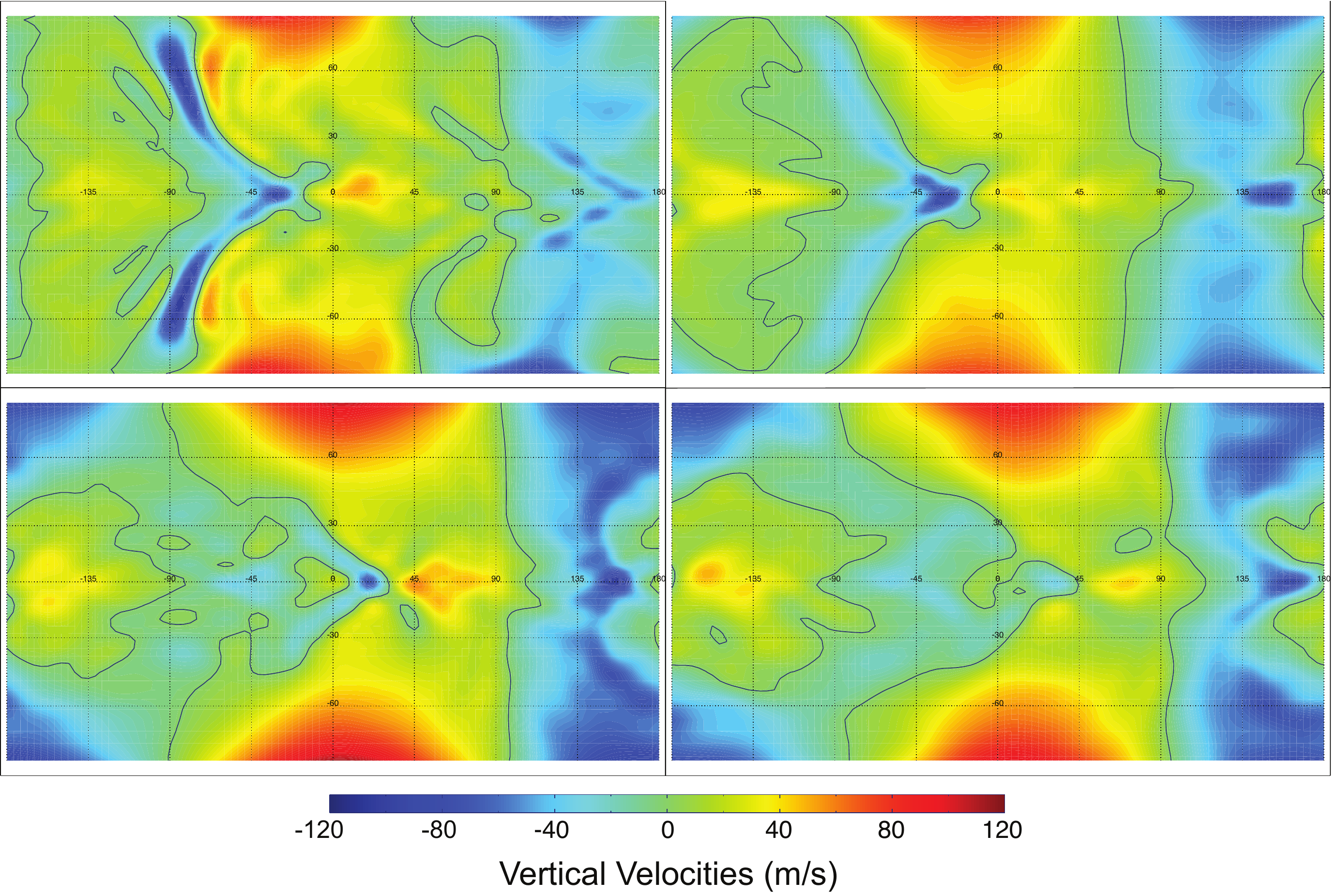}
\centering
\label{fig:vertvel}
\caption{Maps of the vertically velocities, averaged in pressure over a range of heights that include our prescribed aerosol profile, for our four representative cases in Figure 1. Maps are shown in cylindrical projections, centered on the substellar point.  Colors corresponding to positive values mark regions of ascent while negative values indicate descent.  The solid line demarcates zero vertical velocity.   We generally expect that condensates should more easily form in regions of upwelling (positive values). These models use the same scattering parameters listed for Figure 1.}
\end{figure*}
%%%%%%%%%%%%%%%%%%

A dominant, broad eastward jet at equatorial latitudes was common to all the models we investigated.  Zonally averaged, the eastward flow extended down to several bars at the equator, but the latitudinal extent was case dependent (Figure \ref{fig:zonalwinds}). For homogeneous atmospheric models\textemdash namely the clear cases and global cloud case\textemdash the eastward flow essentially extended from pole to pole above the ~5 bar level, with a zonally-averaged westward flow at greater depths. In contrast, for cases with strong, inhomogeneous scattering, the westward flow was present at higher latitudes over most pressures, confining the eastward flow to equatorial and mid-latitudes.  This difference may be explained by differential heating on the dayside of the planet as cloudy longitudes west of the substellar receive significantly less stellar heating compared to longitudes to the east.  This resulting pattern yields westward flow at mid- and high-latitudes for locations west of the substellar longitude.

The precise magnitude and extent of the flows depended only slightly on scattering properties of the aerosols, but only in the western terminator+nightside cloud cases did the choice of mineral prove significant.  For the highest infrared opacity of perovskite, the maximum zonally averaged winds were roughly 15\% greater and peaked higher in the atmosphere compared to the corresponding silica case.

Non-zonal features in the horizontal wind field are evident in the 25 mbar maps shown in Figure \ref{fig:tfield}.  While a strong zonal component is found at all latitudes in the homogenous cases, the pattern changes in the inhomogeneous cloud cases at high latitudes to the west of substellar.  Winds weaken and even reverse in direction around large eddy-like features centered around the western terminator at polar latitudes, disrupting the integrated zonal flow.

Altogether, the salient point is that the strong eastward flow at the western terminator is maintained despite the presence of thick aerosol layers, but only for latitudes less than about 60$^{\circ}$. If aerosols formation were limited to the night side, their advection to the dayside, across the western terminator, would be most efficient at equatorial and mid-latitudes.  Advection at higher latitudes would be limited, and in some locations oppositely directed, by weaker winds.

%%%%%%%%%%%%%%%%%%
% FIGURE 4 W MATRIX
%%%%%%%%%%%%%%%%%%

In addition to the horizontal winds, the vertical velocities may also be used to assess the plausibility of aerosol distributions.  Though clouds of aerosols may form anywhere conditions favor the growth of droplets or particles, persistent aerosols are less favorable in downdrafts (regions of negative vertical velocities) since that setup would require a steady source of material from above, where the atmosphere is generally thinner, colder, and potentially depleted of volatiles.  In contrast, regions of positive vertical velocities may condense and loft particles high into the atmosphere, supplying rich volatile material from below to balance the losses due to evaporation and settling. Settling rates are largely dependent on the size of the aerosol particles, as larger vertical velocities are required to suspend larger particles. As \cite{Parmentier2013} showed for an exoplanetary example, settling can deplete larger aerosols from the visible atmosphere quickly in the absence of strong updrafts; however, based roughly on calculations of \cite{Parmentier2013}, with relatively low gravitational acceleration and sub-micron particle sizes, it is plausible that even modest updrafts (on order of 1-10 m/s) can potentially keep aerosols suspended at low pressures in Kepler 7b's atmosphere.

Since our models assume vertical hydrostatic equilibrium, the vertical velocities are simply the result of applying the continuity equation. Figure 2 shows the mapped vertical velocities averaged from 100 mbar up to 0.3 mbar, spanning a range from below the cloud base to near the top of the cloud layer.

Our modeling suggests that the average vertical velocities throughout the prescribed cloud level are positive over much of the dayside and roughly half the nightside\textemdash from the western terminator to the antistellar longitude.  The regions of upwelling on the nightside beyond the western terminator extends to the poles in the uniform atmospheric models, but are confined to low- and mid-latitudes in the inhomogeneous cloud cases, similar to what was seen with the zonal winds.  Localized intense pockets of updrafts and downdrafts exceeding 80 m s$^{-1}$ coincide with regions of strongly convergent or divergent flow, associated with commonly found chevron-shaped features in the temperature field \citep{RauscherMenou2010}. These features are largely diminished in the inhomogeneous aerosol models suggesting that the pattern of convergence and divergence is partly disrupted by the altered pattern of instellation. Patchy downwelling is found just west of the substellar point, and the greatest area of descent is found is found on the nightside from the eastern terminator to the antistellar point in all cases.  This pattern of upwelling and downwelling would therefore be favorable for aerosols over much of the dayside and portions of the night side (particularly near the equator) from the western terminator to the anti-stellar point, but generally unfavorable for persistent aerosols on the nightside from the eastern terminator to the anti-stellar point.  This conclusion was robust for both minerals (i.e. IR scattering properties) and background albedos investigated.

\subsection{Temperature Field}

%%%%%%%%%%%%%%%%%%

Vertical profiles (Figure 3) and horizontal maps (Figure \ref{fig:tfield}) of the atmospheric temperature structures show how aerosols affect heating.  With the lowest spherical albedo, the clear case unsurprisingly has the highest global averaged temperatures since it absorbs more stellar radiation overall; the stellar radiation also penetrates to greater depths, especially when Rayleigh scattering is neglected, warming pressures between 0.1-10 bars significantly more than in the scattering cases.  Introducing aerosol scattering on the dayside serves to increase the planetary albedo and reduce the global temperatures.  In cloudy regions, our modeled aerosols also serve to reduce the atmospheric transmission in the visible and, given the small particle sizes, to a much lesser extent in the infrared. Consistent with theory and previous modeling \citep[e.g.]{Pierrehumber2010book}, conservative scatterers in the visible produce cooling beneath the aerosols due to attenuation of stellar radiation. This effect dominates over any heating due to infrared scattering on the dayside, as can be seen in vertical profiles along the equator in the western cloud case; however, on the nightside there is no visible scattering and the effect of the infrared scattering dominates.  This nightside infrared scattering warms the layers beneath and within the cloud layer, effectively insulating the atmosphere below.  This can very clearly be seen in nightside (180 $^{\circ}$) temperatures for the western terminator+nightside cloud case in Figure 3 and Figure \ref{fig:tfield} . As expected, the infrared scattering effect is accentuated for the cases using higher infrared optical depths for the aerosols, specifically cases using perovskite\textemdash given its relatively greater infrared extinction efficiency relative to the silica\textemdash and those assuming a greater overall optical thickness\textemdash i.e. cases assuming a greater background albedo.

\begin{figure*}[h!]
\includegraphics[width=\textwidth,keepaspectratio]{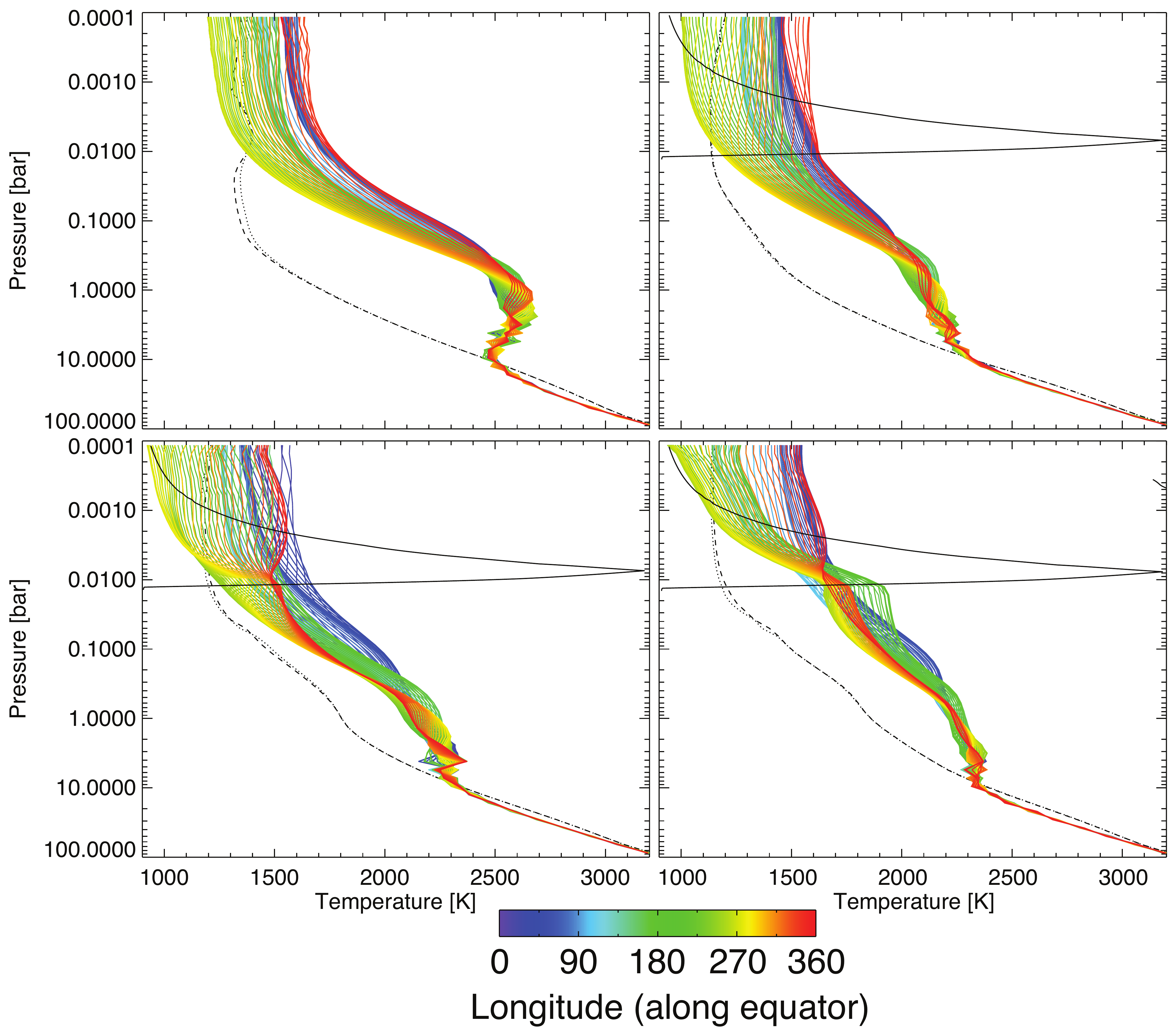}
\centering
\label{fig:tprofs}
\caption{Modeled vertical temperature profiles corresponding to the four representative cases, as in Figure 1. The solid lines are locations along the equator, with the color indicating the longitude (0 $^{\circ}$ is the substellar point).  The dashed and dotted lines are for profiles at the north and south poles. The solid black line depicts the assumed vertical distribution of aerosols, normalized in value to the range of the plot.  In the clear case (upper left panel) the radiation penetrates deepest, producing higher temperatures at most longitudes and at deeper pressures relative to other cases. The global cloud (upper right panel) produces relatively cooler profiles than the clear case, particularly at depth, due to the increased albedo and reduced atmospheric transmission. More complex behavior is seen in the spatially inhomogeneous cases (lower panels). The thick reflective clouds cool the underlying atmosphere west of the substellar point (longitudes 270$^{\circ}$-360$^{\circ}$) in the western terminator case (lower left), while infrared scattering causes warming beneath the clouds, particularly at the anti-stellar point in the western terminator+nightside model (lower right). These models use the same scattering parameters listed for Figure 1.}

\end{figure*}

\begin{figure*}[th!]
\includegraphics[width=\textwidth]{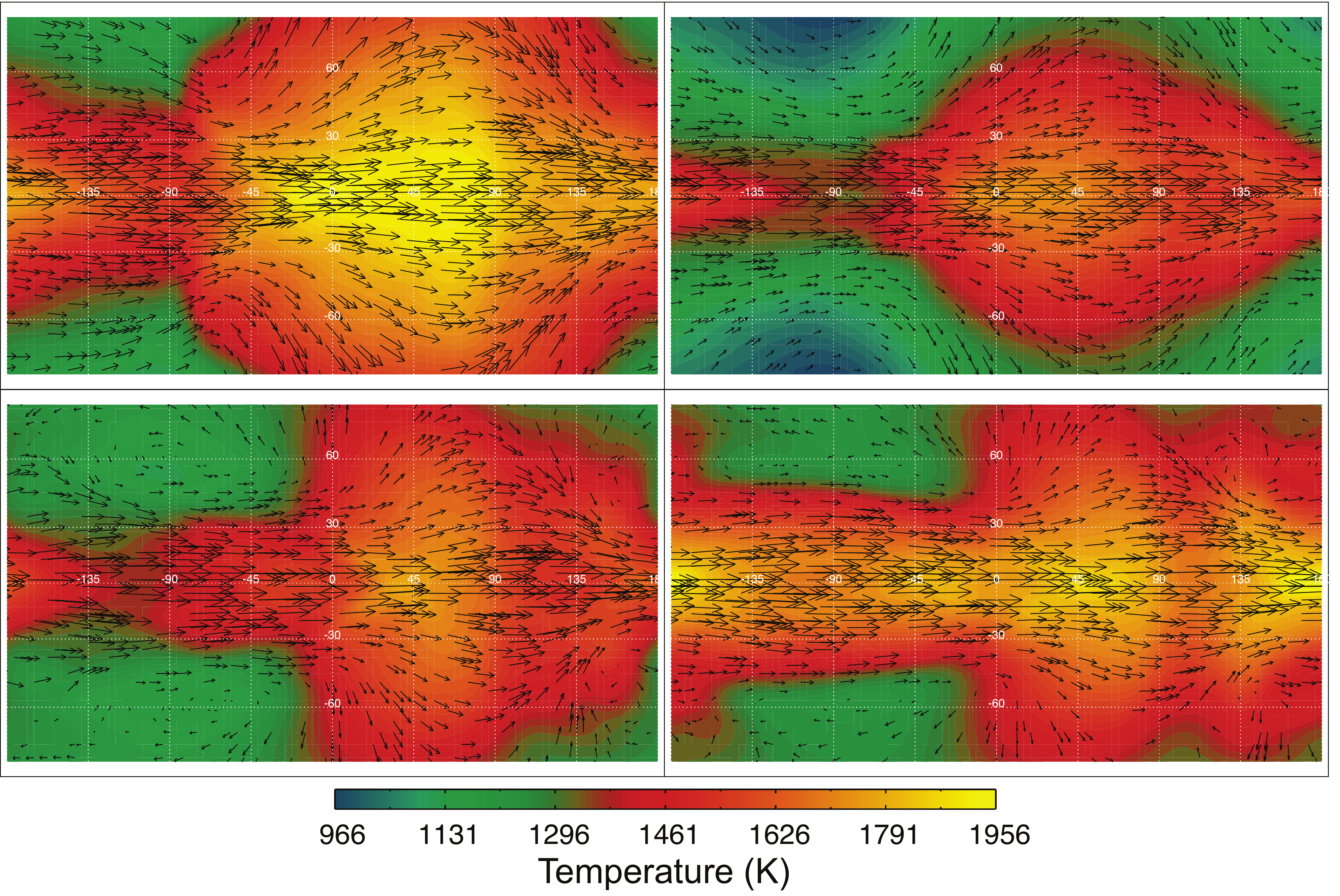}
\caption{Modeled temperature maps for the same representative cases at 25 mbar, near the infrared photosphere for the clear atmosphere and just below the 10 mbar cloud base. Vectors show the direction and relative magnitude of wind field.  The clear and global cloud models show similar temperature patterns, but with different magnitudes due to the difference in spherical albedos, while the inhomogeneous aerosol models show significantly altered structures. The aerosols reduce instellation along the western terminator producing cooling relative to the clear case, but infrared scattering from clouds on the nightside can increase equatorial temperatures below the cloud level at all longitudes. These models use the same scattering parameters listed for Figure 1.}
\centering
\label{fig:tfield}
\end{figure*}
%%%%%%%%%%%%%%%

In both the clear and global cloud cases, the hottest regions on our 25 mbar maps is location east of substellar point, consistent with the offset discussed in other studies \citep{showman2009atmospheric,Dobbs-Dixon2010,RauscherMenou2010}. This is a direct consequence of the relative magnitude of the dynamical and radiative timescales, as irradiated gas is advected by the strong winds \citep{CowanAlgol2011, PerezBeckerShowman2013, KomacekShowman2016,ZhangShowman2017}. The pattern of heating is altered when an inhomogeneous spatial distribution of the scatterers is assumed. For our simple western terminator aerosol model, regions near and to the west of the substellar point are significantly cooler due to the thick, reflective cloud at these longitudes, whereas clear skies allow intense heating the east.  Consequently, the integrated center of the hotspot is shifted further eastward relative to the other cases, producing the greatest shift of all our models. This also results in slightly hotter temperatures over the nightside compared to the global cloud case with equivalent albedo. The nightside warming is greater still for the western terminator+nightside case, in which the atmosphere can transmit heat less readily to space, but it still absorbs the same amount of instellation on the dayside. In order to maintain the same energy balance, the temperatures increases to allow for greater overall emission.  As a result, the western terminator+nightside case has the warmest nightside temperatures and a maximum dayside temperature that rivals the clear case.  The heat retained and advected over the nightside also serves to partly offset the eastward shift in the hot spot, producing a shift that falls between the western terminator case and homogenous atmosphere cases. Differences between the western terminator and western terminator+nightside cases are greatest when the infrared opacity of the clouds is greatest.

\subsection{Outgoing Infrared Radiation and Infrared Phase Curves}

\begin{figure*}[t!]
\includegraphics[width=\textwidth]{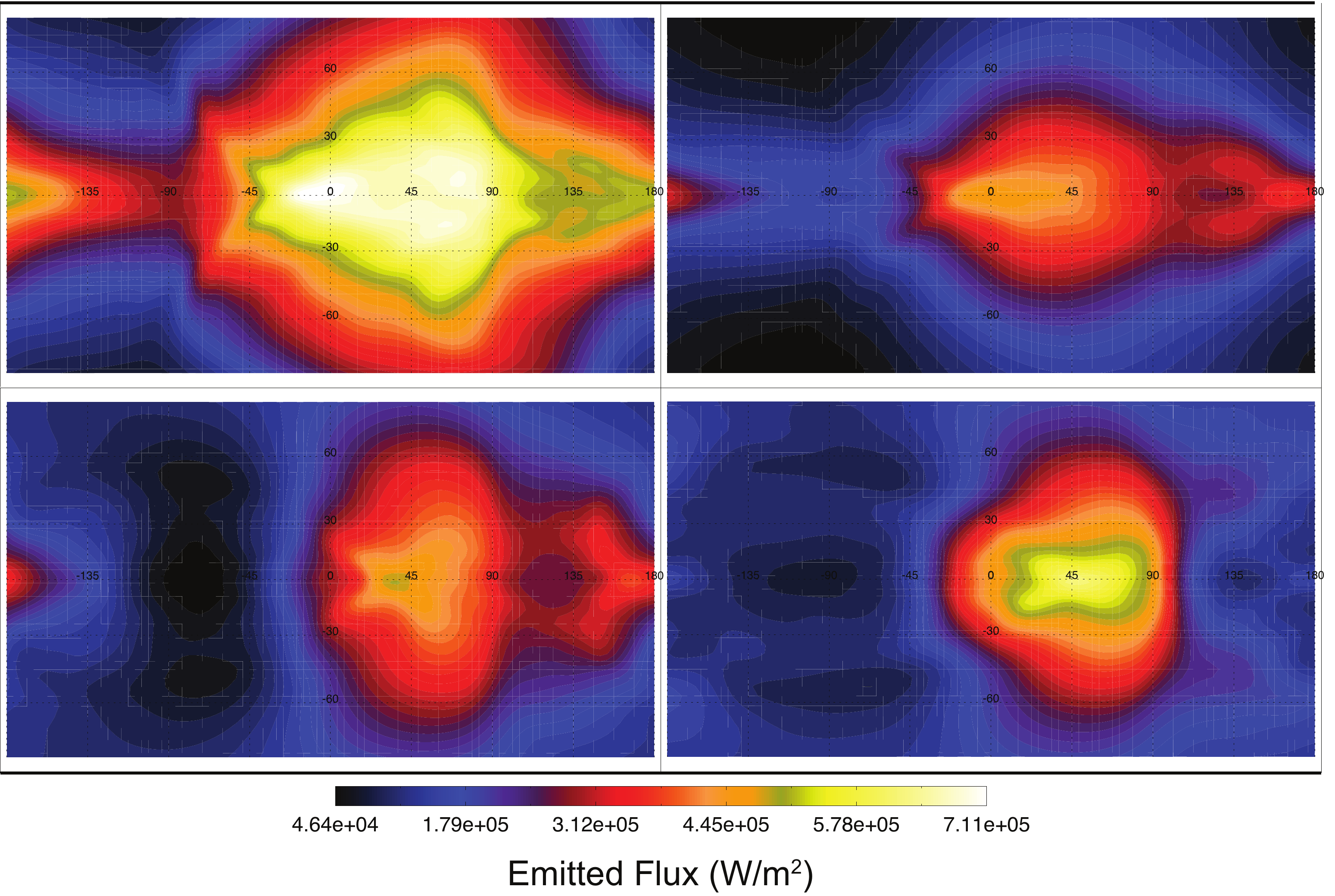}
\caption{Emitted Infrared flux at the top of the model, for our four representative cases, in cylindrical projection centered on the substellar point.  The clear atmosphere (upper left) is the hottest and brightest, but the spatial pattern of emission is fairly similar to the global cloud case (upper right). The flux emitted from the inhomogeneous cloud models (bottom) is significantly altered. The western terminator cloud model (bottom left) and western terminator+nightside model (bottom right) have significantly reduced emission from clouded regions near the western terminator, though the latter also has reduced emission beyond the eastern terminator. These models use the same scattering parameters listed for Figure 1.}
\centering
\label{fig:OLR}
\end{figure*}
%%%%%%%%%%%%%%%%%%

The observed infrared emission is determined by both the temperature field and the infrared transmissivity of the atmosphere. By modifying where radiation is deposited and emitted in the atmosphere, aerosols have the ability to alter this pattern of observed emission.  In Figure \ref{fig:OLR} we present maps of the flux emitted from the top of the planet's atmosphere, for our representative cases.  These are self-consistently produced from the radiative transfer scheme in our GCM.  Integrating these maps of the disk over different observing geometries yields curves of thermal emission from the planet, as a function of orbital phase, as shown in Figure \ref{fig:Phasecurves}.

The greatest emission comes from the clear case, as expected given the higher temperatures and greatest atmospheric transmission. The eastward advection of the hotspot shifts the infrared phase curve to a minimum value prior to transit and to a peak value prior to secondary eclipse, the standard result for hot Jupiters. When scattering is included to increase the global albedo, the emission are significantly reduced at all locations. The corresponding phase curves are mostly shifted in intensity, but there is also a slight shift in phasing. In the western terminator cloud cases, the hotspot is shifted further to the east of the substellar point by the cooling effect of the aerosols to the west; additionally, the aerosols also serve as a source of opacity in the infrared channel, reducing the infrared emission from the underlying atmosphere.  The combined effect significantly reduces the infrared emission west of the substellar point and produces a significant shift to lower phases in the infrared phase curve relative to the other cases. By reducing atmospheric transmission, aerosols might in general be expected to reduce the phase offset by pushing emission to higher levels of the atmosphere where radiative time scales are shorter and the hottest spot is advected less\citep{sudarsky2003}; however, as our modeling shows, this is not always the case.

Interestingly, the western terminator+nightside case has very little shift in phase relative to the clear and global cases.  Furthermore, compared to the western terminator and global cloud cases, the western terminator+nightside case has less emission over much but not all of the nightside, but its minimum value is similar. The aforementioned warming effect serves to partly counter the shift seen in the western terminator case, and the increased aerosol opacity on the nightside is partly compensated by greater emission from hotter gas.  Clouds beyond the eastern terminator greatly suppress emission relative to the other cases, but they also allow heat to be more efficiently transported downwind, contributing to emission beyond the western terminator that exceed emission in the western terminator cloud case.  Partly impeded by clouds elsewhere, heat escapes most efficiently through the clearing on the eastern half of the dayside, yielding a relative greater amplitude peak in the dayside emission compared to the other cases. So though it may be natural to assume that an exoplanet with a cloudy nightside yields significantly less flux from its nightside, in our modeled cases, this is only true for portions of the nightside, and a measure of the absolute minimum flux alone would not constrain the cloud model.

In terms of observational constraints, it is worth noting that attempts to discriminate between cloud models using infrared phase curves could be possible for the brightest hot Jupiters. Kepler-7b is an order of magnitude dimmer in the infrared than the dimmest system for which a full phase curve has been successfully observed, i.e. WASP-19b \citep{Wong2016}, and has a relatively long orbital period ($\sim$4 days), making it a particularly challenging target.  Nonetheless, the conclusions drawn here regarding clear and cloudy nightsides should be seen as potentially applicable to other, more easily observable, hot Jupiter systems.

\begin{figure}[h!]
\includegraphics[width=\columnwidth]{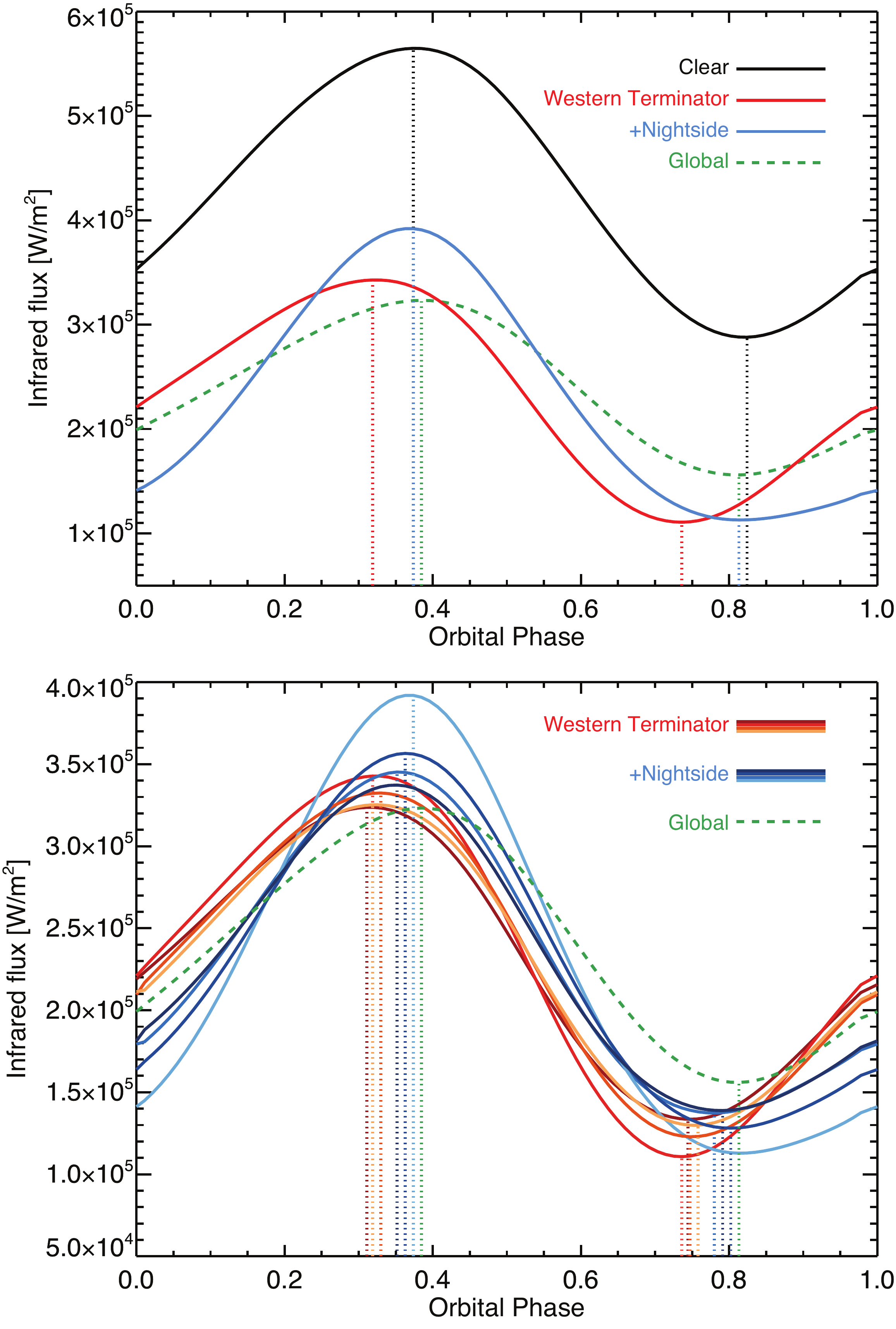}
\centering
\caption{ Modeled infrared phase curves for our four representative cases (top) and all simulations except the clear case (bottom). The black line in the top panel represents the clear case with lower albedo and consequently greater emission. The red line is an example of the western terminator case, the blue is the western terminator+nightside, and the dashed green line is the global cloud case.  The western terminator case has a significant phase shift compared to the others, while clouds on the nightside suppresses emission from phases corresponding to much, but not all of, the nightside. The bottom panel shows these conclusions generally hold for all our assumed infrared scattering and background atmosphere parameters: the assumed inhomogeneous distributions act to shift the peak of the phase to earlier times compared to homogenous models, but the presence of clouds on the nightside reduces this shift.}
\label{fig:Phasecurves}
\end{figure}
%%%%%%%%%%%%%%%%%%

\section{DISCUSSION: Evaluating the Self-consistency of the Aerosol Models}

In the case of Kepler-7b, an aerosol distribution along the western terminator has been proposed and modeled to produce greater reflectance following secondary eclipse \citep{Demory2013,MnI2015}.  Such a distribution may be plausibly explained if temperatures were cool enough to support aerosol growth along the western terminator, or if clouds formed on a cooler nightside and were then advected to the dayside by eastward winds before being vaporized by hotter temperatures \citep{Parmentier2016}.  We now examine whether or not this picture is generally consistent with the results of our dynamical modeling. 

Overlaying the cloud maps on the temperature and wind fields at the chosen 10-mbar cloud base (see Figure \ref{fig:tfieldclouds}), we see that for clouds centered near the western terminator, the temperature pattern is generally consistent with a strong day-night temperature contrast.  In this case, the highly reflective aerosol layer ultimately leads to cooler temperatures near the western terminator, while clear skies radiatively cool the atmosphere on the nightside. If clouds formed on the cooler nightside, strong equatorial winds would blow aerosols from west to east, where they then become warm.  If the dayside is hotter than the condensation temperature, the aerosols would evaporate, and the assumed cloud distribution would be physically plausible.  Based on the temperature maps, evaporation would need to occur around 1200-1400 K at 10 mbar in order to limit clouds to the western terminator and nightside.

%%%%%%%%%%%%%%%%%%

However, with the prescribed clouds in place, the coldest temperatures are not actually limited to the nightside.  Temperatures at high-latitude regions on the dayside are even colder than equatorial regions on the nightside.  If the aerosols are temperature-dependent, these high latitudes would have presumably deeper and thicker aerosols, unlike the assumed distribution, as \cite{Parmentier2016} demonstrated. Furthermore, while equatorial winds may advect material across from the nightside to the dayside at velocities of 5 km/sec, mid-latitude and polar winds would actually advect material in the opposite direction, from dayside to nightside, in the inhomogeneous cloud cases.  Therefore, while aerosols at the equator may be advected from the nightside, aerosols at higher latitudes could not, but temperatures may allow these aerosols to form in place with no need for advection. 

\begin{figure}[h!]
\includegraphics[width=\columnwidth]{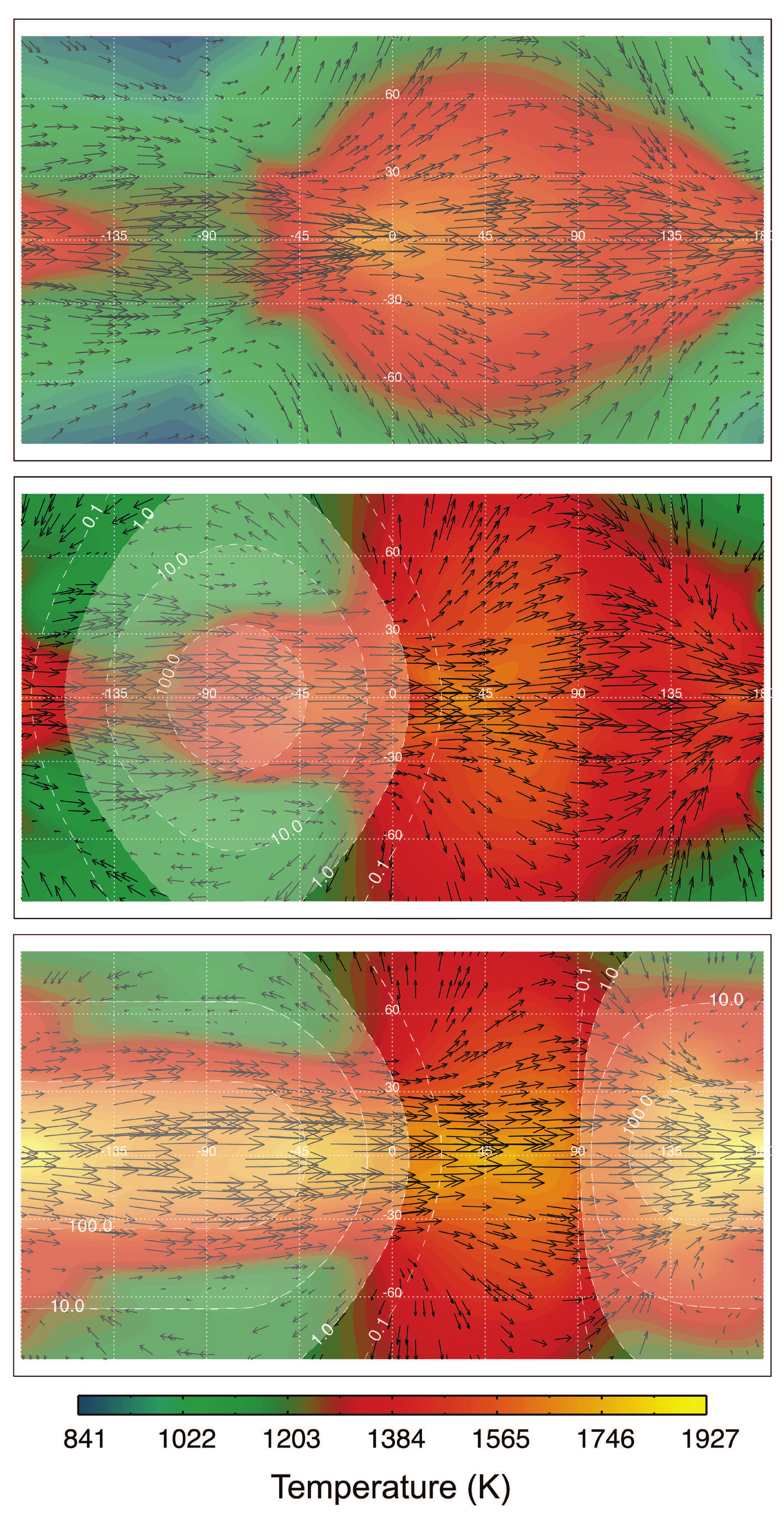}
\centering
\caption{ Aerosol cumulative optical thickness for three cloud models superimposed over the temperature and wind vectors at 10 mbar, the pressure of the cloud base (not the 25 mbar level shown in Figure 4); (top) global cloud model featuring clouds at all locations; (center) western terminator cloud model centered 75$^{\circ}$ west of substellar point based on results of \cite{MnI2015}; and (bottom) our western terminator+nightside cloud model, featuring heavy cloud cover extended to the nightside. Cumulative optical depth contours show $\tau \ge 1$ shaded in white. In this modeling, the prescribed clouds affect the temperature and winds, but do not respond to the environment. These models use the same scattering parameters listed for Figure 1.}
\label{fig:tfieldclouds}
\end{figure}

Finally, considering vertical velocities in the same regions for this cloudy case, on the dayside, upwelling is found at high latitudes, while weak downwelling is present at the equator west of the sub-steller point (see Figure 2). This pattern is reversed on the nightside, where the upwelling is preferentially at the equator and east of the anti-stellar point.  If upwelling is assumed to favor aerosol formation, this pattern is again consistent with formation on the nightside followed by eastward advection at the equator, but with formation in situ at high latitudes.   

The above scheme, however, fails if additional aerosols on the nightside increase the infrared opacity. In these cases, day-night contrast at the equator is significantly diminished due to the retention of heat by the insulating cloud layer.  Parts of the equatorial nightside would be just as hot as much of the dayside, and hence presumably no more favorable for temperature-dependent aerosol growth.  Though prescribing nightside clouds in our simple modeling results in inconsistencies, we wish to stress that this result is for a specific case and does not in any way preclude the presence of clouds on the night side in general.  Our modeling assumes a single infrared passband and a uniform, thick aerosol coverage within each clouded grid point (i.e. a cloud fraction equal to unity).  In reality, condensate clouds may be patchy, allowing more heat to escape through optically thin breaks in the aerosol cover.  As \cite{Lee_kitchensink2016} found, more complex modeling would favor a inhomogeneous, wavelength-dependent cloud opacity with properties differing in longitude, latitude, and depth. Furthermore, our simple aerosol modeling is unresponsive and completely disregards the complex microphysics and climate feedbacks associated with cloud cover.  More sophisticated modeling over a wider parameter space is left for future work, but our simple modeling shows that if the clouds are assumed thick, uniform, and expansive enough to prevent significant infrared cooling to space, the retained heat may be considerable enough to erode some of the cloud deck through evaporation or enhanced mixing; this could potentially place some limits on cloud opacity, and a balance should be obtained by including feedbacks for a self-consistent solution.

\section{SUMMARY}

Motivated by the observational evidence suggesting inhomogeneous clouds in exoplanetary atmospheres, we investigated how proposed simple cloud distributions can affect atmospheric circulations and infrared emission.  To this end, we simulated temperatures and winds for the hot Jupiter Kepler-7b using a GCM that included aerosol modeling.  We assumed fixed aerosol distributions based on the distributions previously inferred from optical phase curves \citep{Demory2013,MnI2015}. These included cases in which aerosols were centered near the western terminator and cases in which this distribution extended across much of the planet's nightside, as well as a clear and global cloud case for comparison.  We investigated how such inhomogeneous cloud distributions affected the atmospheric temperature and wind structure, and whether or not the results were actually consistent with clouds being advected from a cooler nightside. 

For all models, we found that the strong eastward jet capable of advecting aerosols from the nightside (across the western terminator) persisted, but only at equatorial latitudes when an inhomogeneous cloud was assumed. In these cases, winds at higher latitudes were less intense and blew from day to night, inconsistent with nightside origin above $\sim$45$^{\circ}$. Nevertheless, colder temperatures may allow temperature dependent aerosols to form in-situ at high latitudes on the day side near the western terminator, as shown by \cite{Parmentier2016}, thus negating the need for advection at these locations. If aerosols evaporate at temperatures around 1200-1400 K at $\sim$10 mbar, then the temperature gradients present in our models suggest that aerosols should evaporate as they are blown from the nightside to the dayside. This could allow for an aerosol distribution roughly centered near the western terminator as inferred from observed visible phase curves \citep{MnI2015}.

When an optically thick aerosol layer was additionally extended across most of the night side, much of the circulated heat was retained below the cloud layer. As a result, the day-night temperature contrast was reduced at the cloud base, and the assumed aerosol distribution was less self-consistent with the modeled temperature field.  A more complex model with radiative feedbacks would be necessary to produce a more self-consistent cloud model.  Our simple treatment nonetheless suggests that extensive, thick, uniform nightside-limited clouds may be difficult to sustain if temperatures are only marginally cool enough for aerosols to form, and this should be considered when invoking nightside-limited clouds to explain suppressed nightside emission. For hot Jupiters similar to those modeled here, our modeling suggests clouds would be more likely to exist at higher latitudes near and beyond the western terminator.

The computed infrared phase curves also show that it may be possible to observationally differentiate between different aerosol models based on the infrared emission.  A global cloud model can produce a phase shifts similar to that of a purely absorbing case, but at much reduced flux, whereas a planet with an abundance of aerosols along the western terminator can produce greater shift due to asymmetric heating that effectively shifts the hottest spot further east of the sub-stellar point.  If high, optically thick clouds are extended across most of the nightside, the insulating effect can largely counter this shift.  For models of equivalent albedo, the amplitude of the phase curves is least in the global cloud case. Though emission from much, but not all, of the nightside is generally reduced when thick nightside clouds are present, the minimum emission can be comparable to other cloud models and this alone is therefore not a strong indicator of the global cloud distribution.

This modeling has shown how a simple treatment of aerosols, consistent with observations, may affect the general circulation of a hot-Jupiter atmosphere, with potentially observable implications.  In future work, we aim to investigate a greater range of cloud models and explore simple methods for increasing the versatility and realism of our simple aerosol modeling while retaining the benefits of a computationally efficient, uncomplicated model. 

\acknowledgements 
We wish to thank Brian Toon for generously sharing source code that we adapted for our radiative transport calculations. This research was supported by NASA Astrophysics Theory Program grant NNX17AG25G.

\bibliography{mybib} 
\bibliographystyle{apj}

\appendix
\section{Results for different Model and scattering parameters}\label{AppB}
We ran simulations of the winds and temperature fields using various parameters consistent with scattering solutions proposed by \cite{MnI2015}.  While representative cases of each are included in the main text, here we show results for two different albedo assumptions, two choices for the infrared scattering properties (based on silica vs. perovskite), and our two basic cloud distribution (western terminator vs. western terminator + nightside). Table 3 provides the parameter values for each accompanying panel in Figures 8 and 9.  The parameters are as defined in Section 2.2.2 of the text.

The greatest differences between results are due to the presence or absence of aerosols on the night side, seen by comparing the left and right columns. The next most significant factor is the infrared opacity of the aerosol layer.  The second row (panels  2 and 6) of Figures 8 and 9 have aerosols with the greatest infrared opacities, as based on the scattering properties of perovskite grains relative to silica.  This greater opacity results in the greatest temperatures and thermal emission. Following this, the effects of the asymmetry parameter and single scattering albedo are not clearly isolated in this limited sampling, but clearly have less effect on the results. The horizontal spatial distribution dominates, and the effect is greatest when the opacity of the aerosol is greatest.

\begin{deluxetable}{cccccccc}[b!]
 \centering
\tabletypesize{\footnotesize}
\tablecolumns{8} 
\tablewidth{4in}
 \tablecaption{Scattering Parameters for Figures 8 \& 9 in Appendix A
 \label{tab:aptab}}
 \tablehead{
 \colhead{Panel No.,} & \colhead{Model} & \colhead{Wavelength} & \colhead{$\tau$ } & \colhead{$\varpi_{0}$} &\colhead{$g_0$} & \colhead{$r_g$}  &\colhead{$\Delta$$\phi$$_c$}}
 \startdata 
 1 & Western Terminator & Visible & 159 & 1.0 & 0.376 & 0 & -65\\
 {}&{}&IR & 0.95 & 0.045 & 0.006 & 0 & -65 \\
{}&{}&{}&{}&{}&{}&{}&{}\\
  2 & Western Terminator & Visible & 159 & 1.0 & 0.376 & 0 & -65\\
 {}&{}&IR & 3.18 & 0.947 & 0.089 & 0 & -65\\
 {}&{}&{}&{}&{}&{}&{}&{}\\
  3 & Western Terminator & Visible & 254 & 1.0 & 0.416 & 0.1 & -75\\
 {}&{}&IR & 1.23 & 0.061 & 0.007 & 0 & -75 \\
{}&{}&{}&{}&{}&{}&{}&{}\\
  4 & Western Terminator & Visible & 254 & 1.0 & 0.416 & 0.1 & -75\\
 {}&{}&IR & 0.59 & 0.785 & 0.028 & 0 & -75\\
 {}&{}&{}&{}&{}&{}&{}&{}\\
  5 & Western Terminator & Visible & 159 & 1.0 & 0.376 & 0 & -65\\
 {}& +Nightside &IR & 0.95 & 0.045 & 0.006 & 0 & -65 \\
{}&{}&{}&{}&{}&{}&{}&{}\\
  6 & Western Terminator & Visible & 159 & 1.0 & 0.376 & 0 & -65\\
 {}& +Nightside &IR & 3.18 & 0.947 & 0.089 & 0 & -65\\
 {}&{}&{}&{}&{}&{}&{}&{}\\
  7 & Western Terminator & Visible & 254 & 1.0 & 0.416 & 0.1 & -75\\
 {}& +Nightside &IR & 1.23 & 0.061 & 0.007 & 0 & -75 \\
{}&{}&{}&{}&{}&{}&{}&{}\\
  8 & Western Terminator & Visible & 254 & 1.0 & 0.416 & 0.1 & -75\\
 {}& +Nightside &IR & 0.59 & 0.785 & 0.028 & 0 & -75\\
 \enddata
\tablecomments{ $\tau$, $\varpi_0$, and $g_0$ refer to the aerosol component of the maximum optical depth, single scattering albedo, and asymmetry parameter, as defined in the text.}
\end{deluxetable}

\begin{figure}[h!]
\includegraphics[width=\columnwidth]{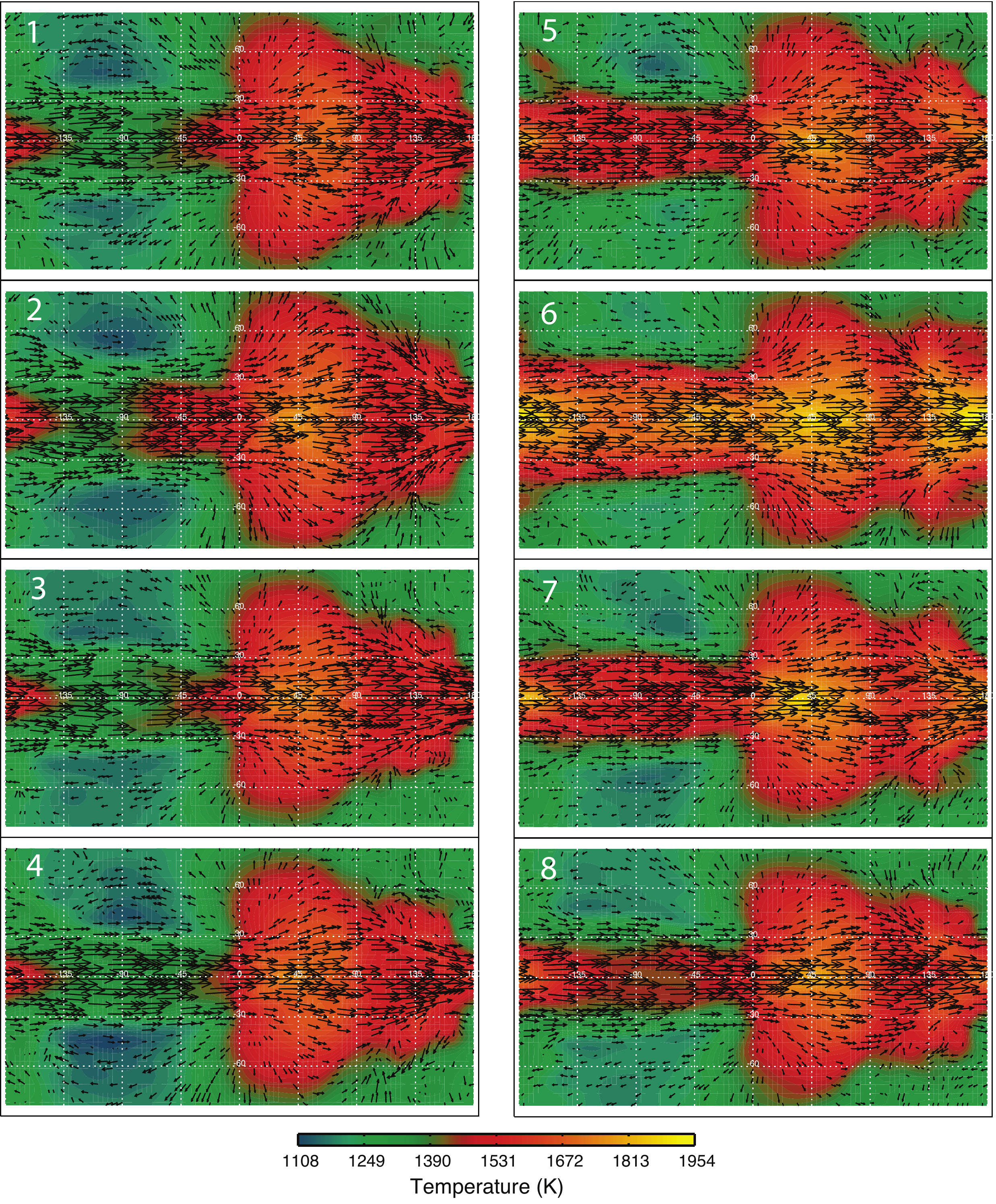}
\centering
\caption{ Modeled temperatures (filled contours) and wind field (vectors)  at 25 mbar, near the infrared photosphere for the clear atmosphere, for the eight cases detailed in Table 3. The western terminator cases are on the left, while the western terminator+nightside cases are on the right.  While the greatest differences between the results can be attributed to aerosol distribution, the choice of scattering parameters has a secondary effect.  The most significant scattering factor is the thermal opacity of the aerosol layer, which is greatest is panels 2 and 6.}

\label{fig:tfield_app}
\end{figure}

\begin{figure}[h!]
\includegraphics[width=\columnwidth]{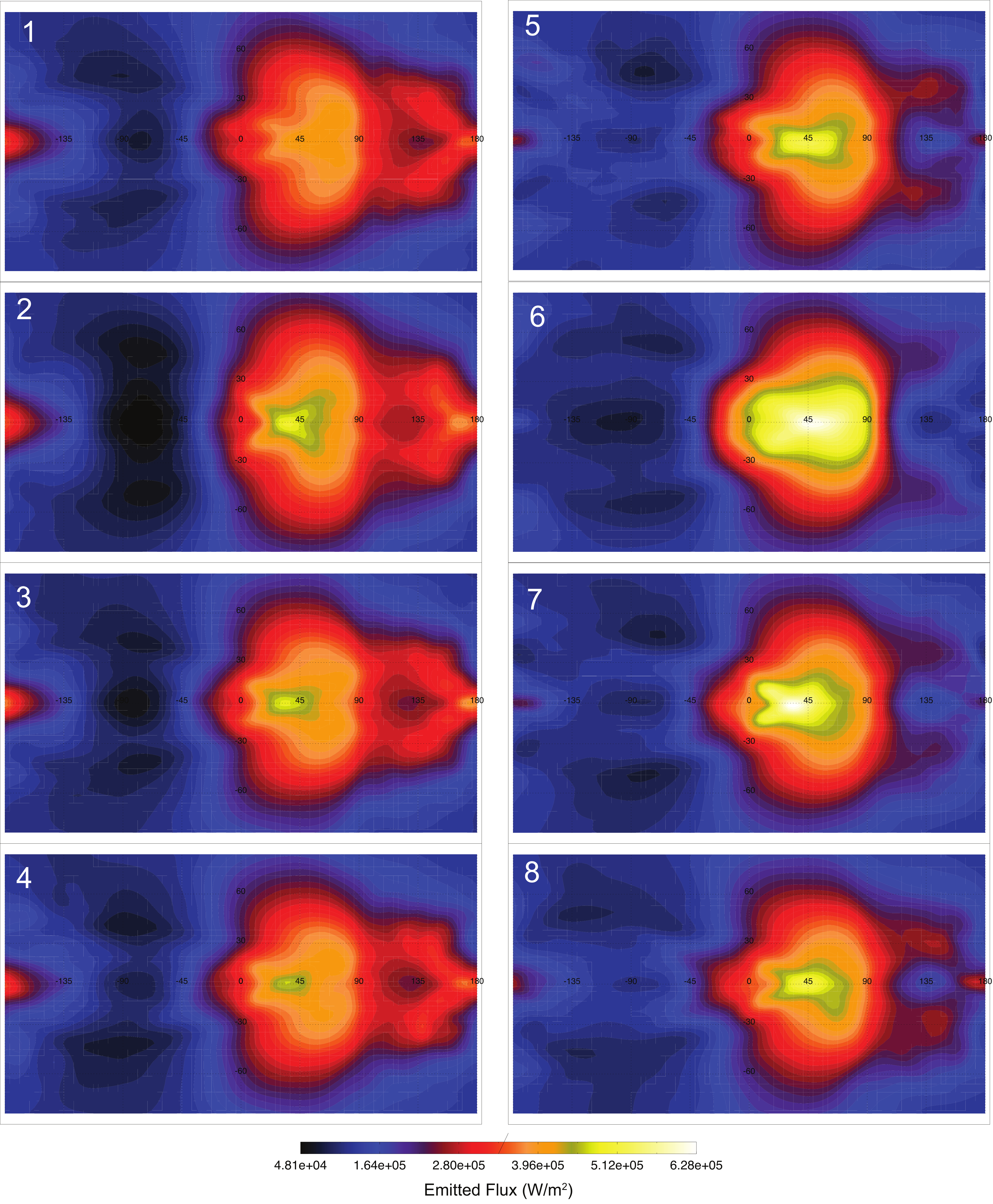}
\centering
\caption{ The thermal emission for the same simulations in Figure 8 and Table 3.  The greatest emissions on the dayside come from simulations with the greatest aerosol thermal opacity (i.e. panels 2 and 6, followed by panels 3 and 7).  The choice of infrared asymmetry parameter and single scattering albedo were less significant and not clearly demonstrated by our limited parameter space.}
\label{fig:olr_app}
\end{figure}

\end{document}